# Two-photon interferences of weak coherent lights

Heonoh Kim[1], Osung Kwon[2,†] & Han Seb Moon[1,*]

[1]Department of Physics, Pusan National University, Geumjeong-Gu, Busan 46241, South Korea

[2]Affiliated Institute of Electronics and Telecommunications Research Institute, Daejeon 34044, South Korea

†E-mail: oskwon@nsr.re.kr

*Corresponding author: hsmoon@pusan.ac.kr

**Abstract**

Multiphoton interference is an important phenomenon in modern quantum mechanics and experimental quantum optics, and it is fundamental for the development of quantum information science and technologies. Over the last three decades, several theoretical and experimental studies have been performed to understand the essential principles underlying such interference and to explore potential applications. Recently, the two-photon interference (TPI) of phase-randomized weak coherent states has played a key role in the realization of long-distance quantum communication based on the use of classical light sources. In this context, we investigated TPI experiments with weak coherent pulses at the single-photon level and quantitatively analyzed the results in terms of the single- and coincidence-counting rates and one- and two-photon interference-fringe shapes. We experimentally examined the Hong-Ou-Mandel-type TPI of phase-randomized weak coherent pulses to compare the TPI effect with that of correlated photons. Further experiments were also performed with two temporally- and spatially separated weak coherent pulses. Although the observed interference results, including the results of visibility and fringe shape, can be suitably explained by classical intensity correlation, the physics underlying the TPI effect needs to be interpreted as the interference between the two-photon states at the single-photon level within the utilized interferometer. The results of this study can provide a more comprehensive understanding of the TPI of coherent light at the single-photon level.

**Introduction**

The observation of two-photon interference (TPI), particularly, the Hong-Ou-Mandel (HOM) effect[1], is fundamental for understanding the superposition principle in quantum mechanics[2], as well as for the development of photonic quantum information technologies such as linear optical quantum computing[3], quantum communication[4], and quantum metrology[5]. In general, the observation of the HOM effect via the superposition of two individual photons at a beam splitter is considered to be a highly reliable method for verifying the indistinguishability of distinct photons[6-11]. Subsequent to the seminal work by Hong, Ou, and Mandel, many TPI experiments have been extensively performed by employing highly correlated photon pairs, to study the fundamental physics underlying two-photon correlations and to explore quantum technologies[12]. The TPI effect is usually interpreted as the interference between the two indistinguishable two-photon amplitudes within the utilized interferometers. Among these, remarkable HOM-type TPI experiments have been performed by employing various types of input states such as two temporally separated photons[13,14], the superposed state in polarizations and frequencies of input photons[15,16], and two photons distributed in spatially separated paths[17].

Meanwhile, HOM-type TPI experiments have been performed with classical light sources such as the weak coherent and fluorescent light beams for measurement of the coherence time and pulse width of ultrafast optical pulses, which are based on the second-order intensity-correlation and photon-coincidence counting techniques[18-20]. Further studies have been performed to demonstrate classical analogue of the HOM effect and to simulate the quantum optical phenomena by using classical light sources[21-25]. More recently, the measurement of high-visibility HOM fringes with weak coherent light at the single photon level has played a key role in the practical implementation of measurement-device-independent quantum key distribution protocols[26-31].

To date, most HOM-type TPI experiments with weak coherent states have been performed using the Mach-Zehnder and polarization-based Michelson interferometers by employing a phase randomization mechanism introduced in one of the interferometer arms[32-38]. In this case, the TPI effect observed via coincidence measurement with two single-photon detectors (SPDs), even at the single-photon level, can also be described classically by intensity correlation. Consequently, the observed visibility bound of 0.5 has been particularly referred to emphasize the classical effect of the HOM-type TPI of coherent light. However, this approach based on intensity correlation does not consider the two-photon amplitudes contributing to the TPI effect at the single photon level. Moreover, when considering the intensity correlation, the observed interference-visibility does not change whatever the intensity involved in the experiment. In practice, the observed TPI-visibility shows the dependence of the mean-photon number in input state[33,34]. Therefore, the physics underlying the interference effect observed even with classical light needs to be understood in the quantum framework as it originate from the interference of the two indistinguishable two-photon amplitudes at the single-photon level.

In this context, in this paper, we report the experimental demonstrations of the TPI of weak coherent pulses. In particular, the HOM-type TPI experiment with phase-randomized weak coherent pulses is examined to compare the TPI effect at the single-photon level with that of correlated photons. We quantitatively analyze the results, including the single- and coincidence-counting rates, based on the statistical properties of the coherent state and the interference fringe originating from the two-photon state at the single-photon level. Moreover, further experiments with two temporally- and spatially separated weak coherent pulses are performed to emphasize the TPI resulting from the two-photon states within the utilized interferometer.

**Results**

**1. Two-photon interference of weak coherent light.**

The coherent state of light is represented by the linear superposition of the photon-number states as

$$|\alpha\rangle = e^{-\frac{|\alpha|^2}{2}} \sum_{n=0}^{\infty} \frac{\alpha^n}{\sqrt{n!}} |n\rangle, \qquad (1)$$

where $n$ denotes the number of photons and $|\alpha|^2 = \langle n \rangle$ the average photon number. Consequently, the probability of $n$ photons being measured within a certain time interval is given by

$$P(n) = \frac{\langle n \rangle^n}{n!} e^{-\langle n \rangle}, \qquad (2)$$

which represents the Poisson distribution. Upon considering the two weak coherent pulses, regardless of whether the two photons originate from two independent sources or from a common source (Fig. 1), the probability of simultaneously finding the number of photons $n_1$ and $n_2$ from the two pulses can be expressed as the product of the two corresponding probabilities. Owing to the statistical property of the coherent state, when the two pulses have the same mean photon number, the probability of simultaneously finding only one photon from each source is equal to the sum of the probabilities of finding two photons from only one source. Thus, we have

$$P(1,1) = P(2,0) + P(0,2) = \langle n \rangle^2 e^{-2\langle n \rangle} \text{ for } \langle n_1 \rangle = \langle n_2 \rangle = \langle n \rangle. \qquad (3)$$

This relation implies that the two methods, shown in Fig. 1, for preparing weak coherent pulses for the TPI experiment are equivalent in terms of photon statistics except for the spectral property.

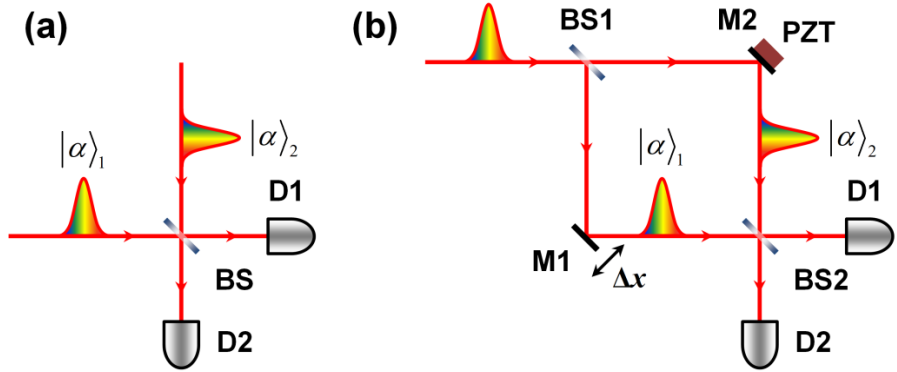

Fig. 1. Two types of experimental schemes for realizing the two-photon interference of weak coherent pulses. The two photons contributing to the interference originate from (a) two *independent* sources or (b) a *common* source. BS, beam splitter; M, mirror; PZT, piezoelectric transducer; D, single-photon detector.

Here, we consider low-intensity coherent light to examine the TPI effect at the single-photon level. For a weak coherent pulse with mean photon number $\langle n \rangle$, the ratio of $P(n)$ to $P(n+1)$ as a function of $\langle n \rangle$ is given by $P(n)/P(n+1) = (n+1)/\langle n \rangle$, which implies that the contribution of more than two photons in the TPI experiment can be ignored for very low values of mean photon number $\langle n \rangle$. Figure 2 shows the statistical property of the weak coherent light: Fig. 2(a) shows photon-number distribution $P(n)$ of the coherent light as a function of the photon number with mean photon number $\langle n \rangle = 0.01$, whereas Fig. 2(b) shows the ratio of $P(n)$ to $P(n+1)$ as a function of $\langle n \rangle$. Under this condition, the ratios of $P(1)/P(2)$ and $P(2)/P(3)$ become 200 and 300, respectively. In our TPI experiment employing coincidence counting with two SPDs, the detection probability of one-photon component $P(1)$ was not considered, although an extremely large number of photons contributed to the single-photon counting events.

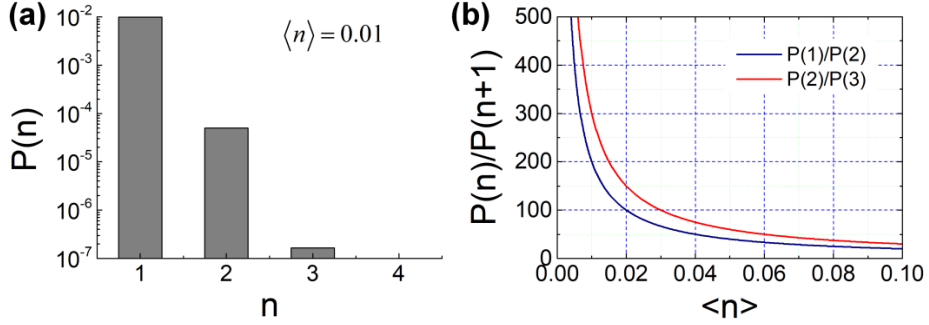

Fig. 2. (a) Photon-number distribution $P(n)$ of coherent light as a function of the number of photons with mean photon number $\langle n \rangle = 0.01$. (b) Ratio of $P(n)$ to $P(n+1)$ as a function of $\langle n \rangle$.

Here, we consider a Mach-Zehnder interferometer (MZI) to perform the TPI experiment with two weak coherent states $|\alpha\rangle_1$ and $|\alpha\rangle_2$, as shown in Fig. 1(b), which corresponds to the standard setup for observing the HOM-type TPI by employing phase-randomized weak coherent light. Considering that the weak coherent pulse in the single-photon level incident on the MZI contains only two photons ($n=2$), two types of two-photon states are generated in the two interferometer arms. One corresponds to the case of two incoming photons traveling through common MZI path, which is similar to the path-entangled state formed with two single photons (or the N00N state with $N=2$, $|2,0\rangle + |0,2\rangle$). The other corresponds to the case in which two photons travel separately through different MZI paths, $|1,1\rangle$. Consequently, the two-photon state within the MZI can be expressed as

$$|\Psi(n=2)\rangle = \frac{1}{2}\left(|2\rangle_1|0\rangle_2 - e^{i2\phi}|0\rangle_1|2\rangle_2\right) + \frac{1}{\sqrt{2}}|1\rangle_1|1\rangle_2, \qquad (4)$$

where the subscripts indicate the two MZI paths and $\phi$ is the relative phase difference between the two paths. As is well known, the first term in Eq. (4) contributes to the highly phase-sensitive TPI similar to the case of the path-entangled two-photon states formed by two single photons. In contrast, the second term indicates phase-insensitive HOM interference. Here, the two photons in the path-correlated state contribute individually to the single-photon interference because each photon interferes only with itself and does not interfere with the other.

The single- and coincidence-counting rates, $N_{D1/D2}$ and $N_{D1\&D2}$, respectively, as functions of the relative path-length difference $\Delta x$ in the MZI are given by[35]

$$N_{D1/D2}(\Delta x) = \frac{1}{2} N_{D1,\,\mathrm{max.}/D2,\,\mathrm{max.}} \left[1 \pm \cos\phi \exp\left(-\frac{1}{2}\frac{\Delta x^2}{\sigma^2}\right)\right], \qquad (5)$$

and

$$N_{D1\&D2}(\Delta x) = N_{D1\&D2, \text{max.}} \left[1 - \frac{1}{2}(1+\cos 2\phi)\exp\left(-\frac{\Delta x^2}{\sigma^2}\right)\right], \quad (6)$$

where $N_{D1, \text{max.}/D2, \text{max.}}$ and $N_{D1\&D2, \text{max.}}$ represent the maximum single and coincidence counting rates, respectively, $\phi = 2\pi\Delta x / \lambda$ the relative phase difference between the two interferometer arms, and $\sigma$ the Gaussian width of the fringe envelope. In particular, the TPI fringe corresponding to coincidence counting in Eq. (6) can be separated into two fringes as $N_{D1\&D2}(\Delta x) = \frac{1}{2} N_{D1\&D2, \text{max.}} \left[1 - \cos 2\phi \exp\left(-\frac{\Delta x^2}{\sigma^2}\right)\right]$ and

$N_{D1\&D2}(\Delta x) = \frac{1}{2} N_{D1\&D2, \text{max.}} \left[1 - \exp\left(-\frac{\Delta x^2}{\sigma^2}\right)\right]$. Here, the former fringe results from the path-correlated state (N00N) and the latter from the two-photon state with two separated photons (HOM) [39].

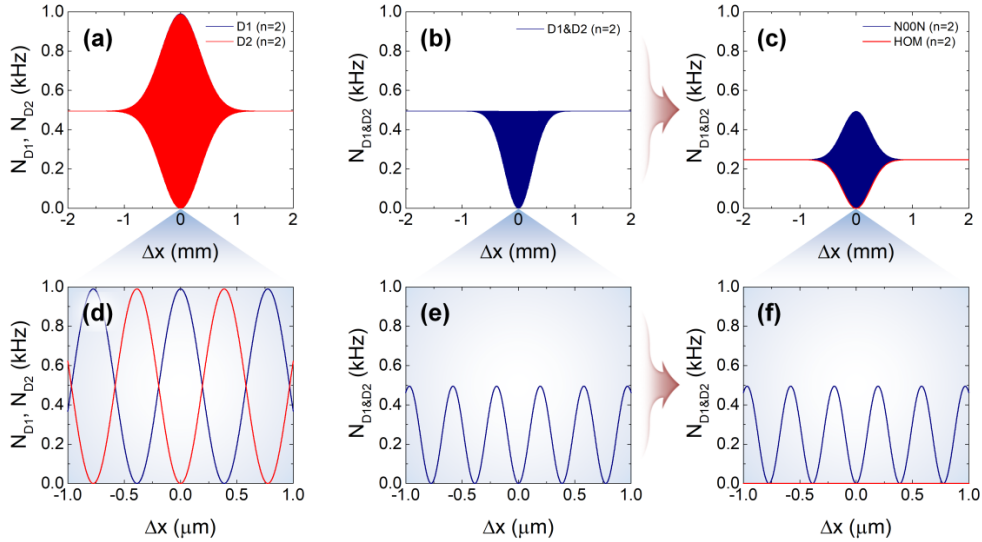

Fig. 3. One- and two-photon interference fringes with weak coherent light with detection probability of two-photon component $P(2)$ for mean photon number $\langle n \rangle = 0.01$. (a) One-photon interference fringes recorded at the two single-photon detectors D1 and D2. (b) Two-photon coincidence fringe composed of the fringes caused by the two two-photon states $|2\rangle_1|0\rangle_2 - e^{i2\phi}|0\rangle_1|2\rangle_2$ (N00N) and $|1\rangle_1|1\rangle_2$ (HOM), as shown in (c). (d,e,f) Interference fringes for $\Delta x \approx 0$.

Figure 3 shows the one- and two-photon interference fringes of weak coherent pulses arising from the two-photon states expressed in Eq. (4). Here, we assumed a coherent pulse with a Gaussian-shaped spectral property ($\sigma = 0.35$ mm) and a center wavelength of 775 nm. From the Poisson distribution with $n = 2$ and $\langle n \rangle = 0.01$ in Eq. (2), when the repetition rate of the weak coherent pulses is 20 MHz, the non-photon-number-resolving SPD at one of the MZI-output ports records the maximum number of counting events of 0.99 kHz, as shown in Figs. 3(a) and 3(d). The TPI fringes shown in Figs. 3(b) and 3(e) include two distinct TPI fringes originating from the two two-photon states expressed in Eq. (4); thus, this fringe can be separated into two coexisting fringes, as shown in Fig. 3(c). Although the two fringes cannot be extracted from a single measurement, these two events do not affect each other.

Here, the registered single- and coincidence-counting rates in Figs. 3(a) and 3(b) originate from the $P(2)$ contribution of only the input pulse for a given mean photon number $\langle n \rangle$. Therefore, the average single and coincidence counting rates at the two SPDs D1 and D2 for the $\Delta x \gg x_{\text{coh.}}$ condition are given by

$$N_{\text{D1/D2}}(\Delta x \gg x_{\text{coh.}}) = N_{\text{D1\&D2}}(\Delta x \gg x_{\text{coh.}}) = \frac{P(2)}{2}f = \frac{\langle n \rangle^2}{4}e^{-\langle n \rangle}f, \qquad (7)$$

where $f$ represents the repetition rate of the weak coherent pulses and $x_{\text{coh.}}$ is the coherence length. However, the two SPDs record photons from both the $P(1)$ and $P(2)$ contributions, and the average single-counting rates for the $\Delta x \gg x_{\text{coh.}}$ condition are given by

$$N_{\text{D1/D2}}(\Delta x \gg x_{\text{coh.}}) = \frac{P(1)+P(2)}{2}f = \frac{1}{4}\left(2\langle n \rangle + \langle n \rangle^2\right)e^{-\langle n \rangle}f, \qquad (8)$$

From the Poisson distribution in Eq. (2), we obtain $P(1) = 9.9 \times 10^{-3}$ and $P(2) = 4.95 \times 10^{-5}$ for $\langle n \rangle = 0.01$; therefore, the maximally observable single- and coincidence-counting rates for $f = 20$ MHz are calculated to be $N_{\text{D1, max./D2, max.}} \sim 198.99$ kHz and $N_{\text{D1\&D2, max.}} \sim 0.495$ kHz, respectively.

Next, we consider a TPI experiment with phase-randomized weak coherent pulses. In this case, the relative phase relation between the two paths of the MZI in Fig. 1(b) needs to be randomized within a short length range by using a phase modulator or piezoelectric transducer (PZT) mounted on mirror M2. Consequently, the terms corresponding to the phase-sensitive interference fringes do not appear in Eqs. (5) and (6). Thus, the phase-insensitive HOM-type TPI fringe revealed only by coincidence counting can be expressed as

$$N_{\text{D1\&D2}}(\Delta x) = N_{\text{D1\&D2, max.}}\left[1 - \frac{1}{2}\exp\left(-\frac{\Delta x^2}{\sigma^2}\right)\right]. \qquad (9)$$

This equation indicates that the maximally observable TPI fringe visibility is limited to 0.5, because the two photons in the phase-sensitive path-correlated state are randomly divided between the two output ports of the MZI and consequently contribute to a constant coincidence regardless of the path-length difference. It is noteworthy that the interference effect and the fringe shape due to the $|1\rangle_1|1\rangle_2$ state in Eq. (4) are identical to those of conventional HOM experiments with two time-correlated single photons, as shown in Fig. 3(c).

## 2. Two-photon interference experiment with weak coherent pulses

We next consider weak coherent pulses at the single-photon level, wherein each pulse does not include more than two photons because the multiphoton contributions to the TPI fringe visibility can be effectively ignored under the experimental condition of a very low mean photon number $\langle n \rangle \ll 1$. As mentioned above, the one-photon events per input pulse are not considered in the two-photon coincidence count, although a large number of photons contribute to the single-counting rates in the experiment.

Figure 4 shows the experimental setup used to observe the one-photon interference (OPI) and TPI fringes of weak coherent pulses. Single-mode fiber (SMF)-coupled weak coherent pulse trains are injected into the polarization-based Michelson interferometer shown in Fig. 4, which is equivalent to the MZI setup shown in Fig. 1(b); however, the Michelson interferometer has an experimental facility for optical alignments to achieve spatial-mode overlap. Pulse-mode coherent light is generated in a mode-locked fiber laser with a 3.5 ps pulse duration at a 775 nm center wavelength and a 20 MHz repetition rate. The laser pulses are highly attenuated to the single-photon level by means of a variable neutral-density filter (VNDF) and subsequently coupled into a single-mode fiber. The interference filter (775-nm center-wavelength and 1-nm bandwidth) and linear polarizer are used to define the spectral and polarization properties, respectively. Two interfering coherent pulses are prepared using a half-wave plate (H1) with its axis oriented at 22.5° followed by a polarizing beam splitter (PBS1). Two quarter-wave plates (Qs) with their axes oriented at 45° are placed in the two interferometer arms to rotate the polarization direction. Thus, the two spatial modes (1 and 2) in Eq. (6) are defined based on the two polarization directions in this experiment employing the polarization-based Michelson interferometer. The second half-wave plate (H2) with its axis oriented at 22.5°, and PBS2 play the role of BS2 of the MZI shown in Fig. 1(b). Two output photons are coupled to the SMF via coupling optics FC1 and FC2 and finally detected by SPDs D1 and D2.

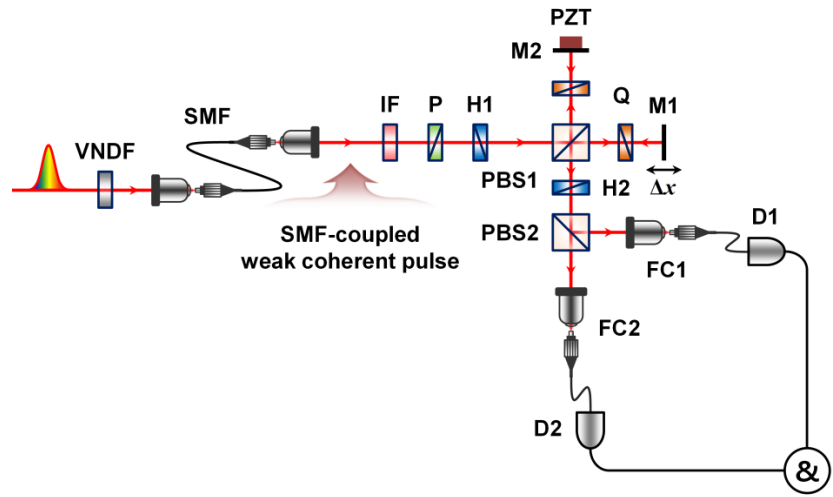

Fig. 4. Experimental setup to observe the two-photon interference of weak coherent pulses. VNDF, variable neutral density filter; IF, interference filter; P, linear polarizer; H, half-wave plate; Q, quarter-wave plate; PBS, polarizing beam splitter; M, mirror; PZT, piezoelectric transducer; FC, Single-mode fiber coupler; D, single-photon detector; & coincidence counting circuit. $\Delta x$ refers to the path-length difference between the two interferometer arms.

First, we examine the interference fringes and counting rates (Fig. 3). In our setup, to measure the OPI and TPI fringes of weak coherent pulses, path-length difference $\Delta x$ between the two interferometer arms was adjusted by moving mirror M1, which was mounted on a motorized translation stage. Figure 5 shows the measured OPI and TPI fringes as a function of path-length difference $\Delta x$. Upon adjusting the VNDF, we estimated the average single ($N_{D1}$ and $N_{D2}$)- and coincidence ($N_{D1\&D2}$)-counting rates at SPDs D1 and D2 as ~300 kHz and ~4.5 kHz (Figs. 5(a) and 5(b), respectively) for a large path-length-mismatch condition ($\Delta x \gg x_{coh.}$). Here, the measured single-counting rates in Figs. 5(a) and 5(c) originate from both $P(1)$ and $P(2)$ of the input pulse for a given mean photon number $\langle n \rangle$. In contrast, the coincidence-counting rate in Figs. 5(b) and 5(d) can originate only from $P(2)$. Assuming ideal conditions (lossless optical system and SPDs with unity detection efficiency), $\langle n \rangle$ is estimated to be ~0.0305. Moreover, for weak coherent light, all coincidence-counting events are caused by accidental coincidences within the resolving time[35]. Therefore, the coincidence can be straightforwardly calculated using two single-counting events as $N_{D1\&D2} = N_{D1}N_{D2}/f$. Here, the single-counting events ($N_{D1}$ and $N_{D2}$) include the photons that originate from both the $P(1)$ and $P(2)$ contributions of the input pulse. The fringe visibilities observed in the single-counting rates are 0.94 ± 0.01 and 0.99 ± 0.01, as estimated from the observed fringes at $\Delta x \approx 0$, as shown in Fig. 5(c). The full-width at half maximum (FWHM) value of the fringe envelope is estimated to be ~0.94 mm, which is determined by the interference filter used in the experiment.

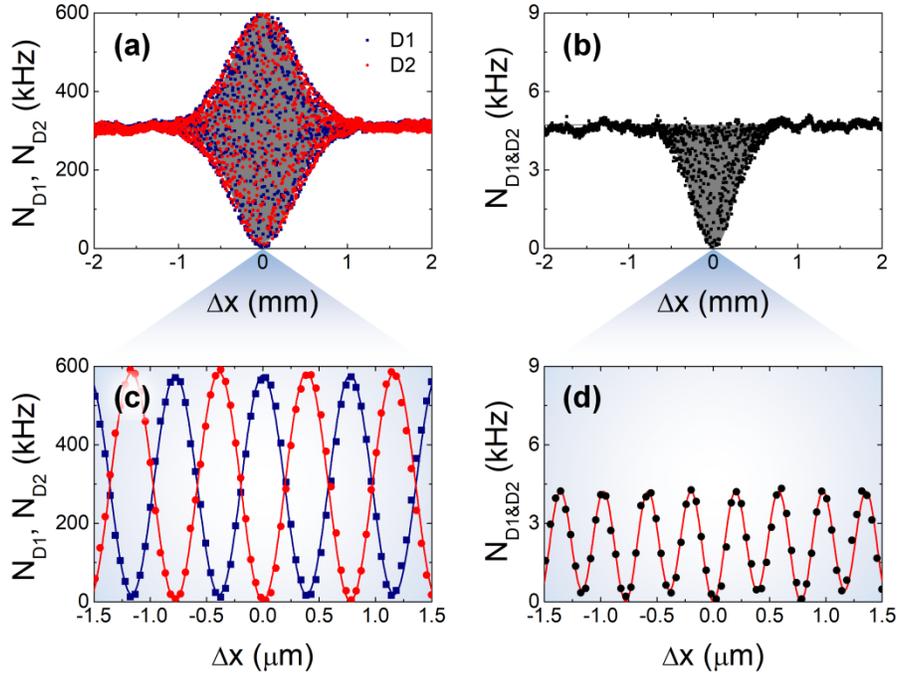

Fig. 5. One- and two-photon interference fringes of weak coherent pulses. (a) Single ($N_{D1}$, $N_{D2}$)- and (b) coincidence ($N_{D1\&D2}$)-counting rates as functions of path-length difference $\Delta x$ (step size of 2 μm). (c,d) Interference fringes measured at $\Delta x \approx 0$.

Meanwhile, the TPI fringe is simultaneously observed in coincidence counting, as shown in Fig. 5(b). As mentioned above, the observed TPI fringe includes two kinds of TPI fringes: the N00N-state fringe due to the path-correlated two-photon state and the HOM fringe due to the two photons separated along the two interferometer arms, as indicated in Eq. (4) and Fig. 3(c). For the exact estimation of the TPI fringe shape and visibility, it is necessary to recall that the TPI fringe of weak coherent light is fully expressed by the accidental coincidence-counting rates as a function of the path-length difference[35], $N_{D1}(\Delta x) \times N_{D2}(\Delta x)/f$. Therefore, the TPI fringe in Eq. (6) can be expressed as

$$N_{D1\&D2}(\Delta x) = N_{D1\&D2,\,max.} \left[ 1 + (V_1 - V_2)\cos\phi \exp\left(-\frac{1}{2}\frac{\Delta x^2}{\sigma^2}\right) - \frac{1}{2}V_1V_2(1+\cos 2\phi)\exp\left(-\frac{\Delta x^2}{\sigma^2}\right) \right], \quad (10)$$

where $V_1$ and $V_2$ represent the fringe visibilities observed for the single-counting rates, as shown in Fig. 5(c). For $V_1 = V_2$, Eq. (10) reduces to Eq. (6). From the theoretical fit to the data points in Fig. 5(d), $V_1$ and $V_2$ were found to be 0.95±0.01 and 0.99±0.01, respectively, which agree well with the values in Fig. 5(c). This result clearly shows that the TPI of weak coherent light should be interpreted by means of the two-photon state in the single-photon picture rather than the classical intensity correlation. The FWHM of the TPI-fringe envelope was estimated to be ~0.58 mm, which is $1/\sqrt{2}$ times narrower than that of the OPI fringe.

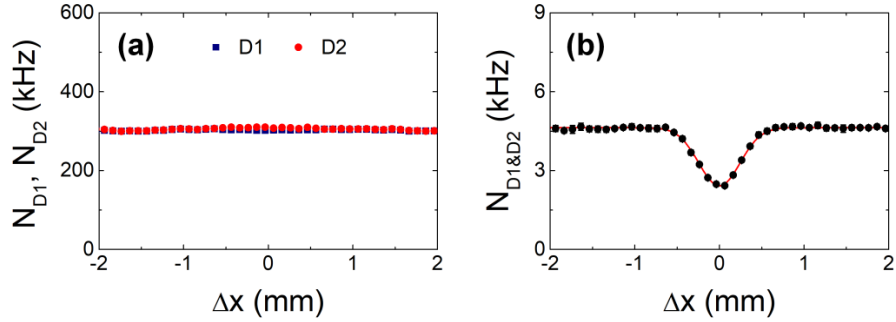

Fig.6. (a) Single ($N_{D1}$, $N_{D2}$)- and (b) coincidence ($N_{D1\&D2}$)-counting rates as functions of the path-length difference when the interferometer arms are phase-randomized.

Next, we examined the case in which the relative phase between the two weak coherent pulses was randomized. To observe the phase-insensitive HOM-type TPI fringe, path-length difference $\Delta x$ is introduced by moving mirror M1, while mirror M2 is affixed to the PZT actuator to randomize the relative phase between the two paths. Phase randomization was originally introduced to ensure that the experimental setup in Fig. 1(b) is equivalent to that in Fig. 1(a), which can effectively make the phase-sensitive interference of two photons in the path-correlated state disappear in the single-and coincidence-counting rates shown in Figs. 3(a) and 3(c), respectively. Therefore, the oscillatory fringes are invisible in single and coincidence counts, as shown in Figs. 5(a) and 5(c), as well as the coincidence counts in Figs. 5(b) and 5(d). Figure 6 shows the experimental TPI results of the phase-randomized weak coherent pulses. The HOM-dip-like TPI fringe with only the $|1\rangle_1|1\rangle_2$ state is shown in Fig. 6(b). From the theoretical fitting, the coincidence-dip fringe visibility and FWHM are found to be 0.48 ± 0.01 and 0.57 ± 0.01 mm, respectively.

## 3. Two-photon interference of temporally separated weak coherent pulses in a phase-randomized interferometer.

Nest we consider two temporally well-separated weak coherent pulses at the single-photon level, wherein each pulse does not include more than two photons. When two such sequential weak coherent pulses are incident on the MZI shown in Fig. 1(b) or the polarization-based Michelson interferometer shown in Fig. 4, a pairwise two-photon state contributing to the HOM effect can be generated within the two interferometer arms. Here, we ignore the case in which two sequential pulses traverse the same path because this two-photon state does not contribute to the TPI in a phase-randomized interferometer. In this regard, it has been recently demonstrated that the pairwise two-photon state with two temporally separated weak coherent photons yields the same HOM fringe as that with the conventional two-photon state[32]. In this work, we performed a HOM-type TPI experiment by employing temporal post-selection by adjusting the coincidence time window in comparison with the temporal separation between sequential pulses. For this purpose, two sequential weak coherent pulses as input two-photon states were prepared with orthogonal polarizations, $|\Psi\rangle = \hat{a}_H^\dagger \hat{a}_V^\dagger (\Delta t)|\alpha_H, \alpha_V\rangle$.

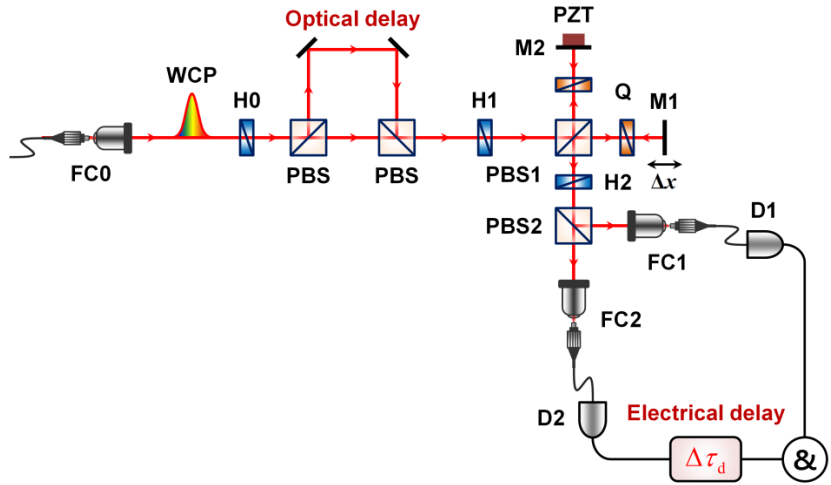

Fig. 7. Experimental setup to observe the two-photon interference of temporally separated weak coherent pulses. Temporal separation between two sequential pulses is introduced by applying an optical delay, and the temporal post-selection of the interfering two-photon states is performed by the application of an electrical delay ($\Delta\tau_d$) and a variable coincidence time window. $\Delta x$ is used for varying the path-length difference between the two interferometer arms to observe interference fringes.

Figure 7 shows the experimental setup in this case. Two temporally separated weak coherent pulses are prepared before their incidence on the phase-randomized interferometer by using an optical delay line, and the selective coincidence counting of the separated pulses is performed by introducing an electrical delay line after one of the SPDs. In our experiment, the optical delay time ($\Delta t$) between sequential pulses was fixed at 8 ns, which corresponds to an optical delay of 2.4 m in free space. The electrical delay time ($\Delta \tau_d$) and coincidence time window ($T_R$) were set to (0 ns, 8 ns) and (4 ns, 10 ns), respectively, for the post-selection of the interfering two-photon states. The pairwise two-photon state related to the HOM interference is generated through the two optical paths of the polarization-based Michelson interferometer as $\hat{a}_H^\dagger \hat{a}_V^\dagger (\Delta t) \to 1/\sqrt{2}\left[\hat{a}_{1,H}^\dagger \hat{a}_{2,V}^\dagger (\Delta t) - \hat{a}_{2,V}^\dagger \hat{a}_{1,H}^\dagger (\Delta t)\right]$. The operating principle of the interferometer is the same as that in Fig. 4, except for the temporal separation between the two photons contributing to the TPI.

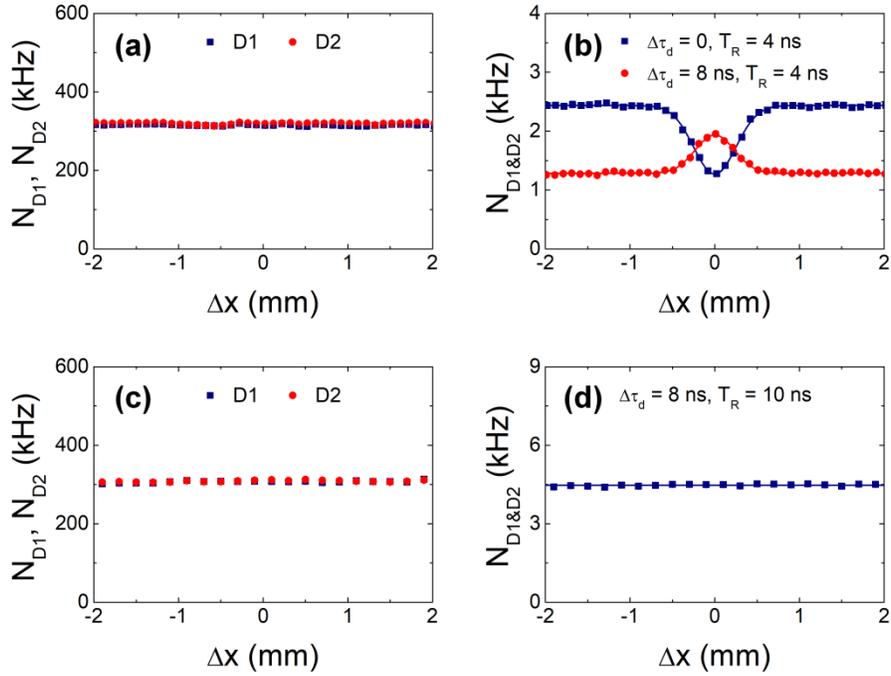

Fig. 8. Single ($N_{D1}$, $N_{D2}$)- and coincidence ($N_{D1\&D2}$)-counting rates as functions of the path-length difference for two temporally separated input photons ($\Delta t = 8$ ns). (a,b) Measurements were performed with ($\Delta \tau_d = 8$ ns) and without ($\Delta \tau_d = 0$) temporal filtering for a coincidence resolving time window of $T_R = 4$ ns. (c,d) Measurements were performed with $\Delta \tau_d = 8$ ns and $T_R = 10$ ns.

Figure 8 shows the experimental results for single- and coincidence-counting rates as functions of path-length difference $\Delta x$. For $\Delta \tau_d = 0$ and $T_R = 4$ ns, the two SPDs measure two photons in the same pulse, and the coincidence-counting circuit records the TPI fringes caused by each pulse. Therefore, the same TPI fringe as in Fig.

6(b) is observed; however, but the coincidence-counting rate is decreased by half for the same condition of the single-counting rates, as shown in Figs. 8(a) and 8(b). This is because the corresponding coincidence-counting rate is given by $N_{D1\&D2}(\Delta x \gg x_{coh.}) = P'(2)/2 \times 2f$, where $P'(2) = (\langle n \rangle/2)^2/2 \times e^{-\langle n \rangle/2}$. In contrast, for $\Delta\tau_d = 8$ ns and $T_R = 4$ ns, the TPI fringe of the pairwise two-photon state can be observed by temporal post-selection in the coincidence-counting circuit. Under this condition, each SPD detects only one photon in each pulse, and the coincidence-counting rate is decreased by one quarter. In this case, the coincidence-counting event arises from $P'(1,1(\Delta t)) = P'(1)P'(1(\Delta t))$, where $P'(1) = P'(1(\Delta t)) = (\langle n \rangle/2) \times e^{-\langle n \rangle/2}$ as expressed in Eq. (3). Therefore, the corresponding coincidence-counting rate is given by $N_{D1\&D2}(\Delta x \gg x_{coh.}) = P'(1)P'(1(\Delta t))/4 \times f$, and the HOM fringe shows coincidence-peak pattern instead of a dip, because the two photons traversing the two interferometer arms are distinguishable in terms of polarization in a given temporal mode. To confirm the temporal post-selection for the observation of the TPI fringe of the pairwise two-photon state, we chose a wider coincidence-counting window than the temporal separation of the sequential pulses ($T_R = 10$ ns). In this case, the coincidence-counting circuit records the two TPI fringes simultaneously, as shown in Fig. 8(b). The resulting coincidence is given by $2P'(2)f$ or $(\langle n \rangle/2)^2 e^{-\langle n \rangle/2} f$, and the interference fringe does not vary with the path-length difference.

## 4. Two-photon interference of spatially separated weak coherent pulses in two phase-randomized interferometers.

Figure 9 shows the experimental setup used to observe the HOM-type TPI of two spatially separated weak coherent pulses, which were prepared using a beam splitter (BS0) followed by a pair of balanced MZIs. In our experiment, we considered the case where only one photon traverses each MZI to show that the TPI effect originates from the two-photon state at the single-photon level and, consequently, is observed via coincidence detection by SPDs D1 and D3 and D2 and D3. Here, the phase shifter (PS) is a thin glass plate with a thickness of ~0.2 mm, which is used to introduce a small optical path delay in one of the four interferometer arms. The two PZTs are used to eliminate the OPI corresponding to the single-counting rate by introducing phase randomization. The two output photons from the interferometers are coupled to SMF couplers and finally detected by SPDs to register single and coincidence counts in the counting electronics.

In this experiment, the necessary conditions for the observation of the HOM-type fringe are the synchronization of the phase randomization and simultaneous change in path-length differences $\Delta x_1$ and $\Delta x_2$ in the two interferometers. In actual experiments, to realize these technical requirements, the two balanced MZIs are constructed of a single polarization-based Michelson interferometer, in which the two optical paths for the spatially separated individual interferometers share common optical components, such as mirrors, wave plates, and polarizing beam splitters.

The experimental results demonstrating the HOM-type TPI effect observed with the two spatially separated interferometers are shown in Fig. 10. The two data sets corresponding to each two-fold coincidence counting with two detectors D1 and D3 and D2 and D3, are obtained simultaneously. During the experiment, the single-counting rates in all the SPDs were maintained constant with varying path-length differences. In the figures, the filled squares, diamonds, and circles correspond to the phase shifts of $\Delta\phi = 0$, $\Delta\phi = \pi/2$, and $\Delta\phi = \pi$, respectively, and the solid lines represent the theoretical curve fits. From the curve fitting, the HOM-fringe visibilities were obtained to be $V_{13}$ = 0.48 ± 0.01 and $V_{23}$ = 0.48 ± 0.01 for $\Delta\phi = 0$ and $\Delta\phi = \pi$, respectively. For $\Delta\phi = \pi/2$, we could not observe any fringe pattern as the path length was varied. As shown in Fig. 10, there is a transition between the HOM peak and dip fringes according to the relative phase between the two amplitudes. For $\Delta x = 0$ in Fig. 10(a), the coincidence counting rate as a function of the rotation angle ($\theta$) of the thin glass plate is given by $N_{D1\&D3}(\theta) = N_{D1\&D3,\infty}\{1+V_{13}\cos[2\pi/\lambda(d-t)]\}$, where $N_{D1\&D3,\infty}$ denotes the coincidences for $\Delta x \gg x_{coh.}$ and $\lambda$ the center wavelength (see the insets in Figs. 10(a) and 10(b)). The increase in path-length difference $d$ with an increase in $\theta$ can be expressed as $d = t/\cos[\sin^{-1}(n_0 \sin\theta/n)]$, where $t$ denotes the thickness of the glass and $n_0$ ($n$) is the refractive index of air (glass).

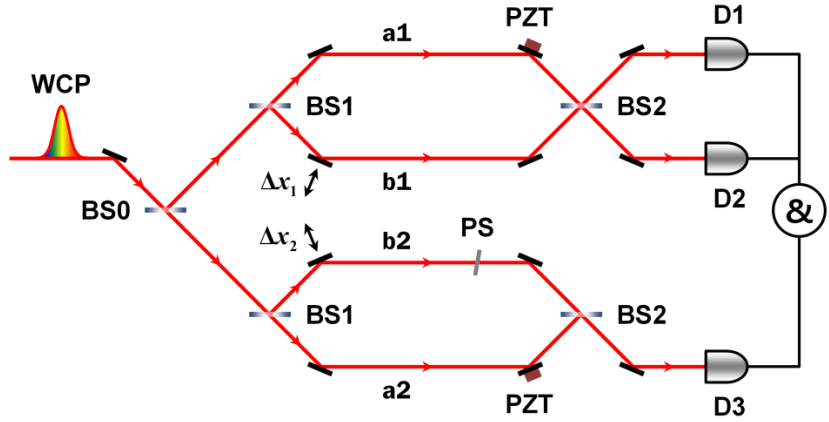

Fig. 9. Experimental setup to observe the two-photon interference of spatially separated weak coherent pulses (WCPs). Two balanced Mach-Zehnder interferometers are positioned after a beam splitter (BS0). BS, 50:50 beam splitter; PS, phase shifter; D, single-photon detector. Phase randomization is synchronously performed by the two PZTs, and $\Delta x_1$ and $\Delta x_2$ simultaneously introduce path-length differences between the two interferometer arms in the two spatially separated interferometers.

To comprehensively understand the experimental results, it is helpful to consider the two-photon amplitudes in both the spatially separated MZIs and the conventional single MZI shown in Fig. 1(b). In our experiment, it is assumed that the phase randomization in the two MZIs is actively synchronized by the two PZTs and path-length differences $\Delta x_1$ and $\Delta x_2$ that are simultaneously introduced as $\Delta x = |\Delta x_1 + \Delta x_2|$, with $|\Delta x_1| = |\Delta x_2|$. In this case, the two interfering two-photon amplitudes, contributing to the HOM-type TPI, can be considered as two pairs of

amplitudes along four separated paths, which correspond to $\hat{a}_{a1}^\dagger \hat{a}_{b2}^\dagger \xrightarrow{BS2} \hat{a}_{D1}^\dagger \hat{a}_{D3}^\dagger$ and $\hat{a}_{a2}^\dagger \hat{a}_{b1}^\dagger \xrightarrow{BS2} \hat{a}_{D1}^\dagger \hat{a}_{D3}^\dagger$. We note here that the coincidence events by the two-photon amplitudes $\hat{a}_{a1}^\dagger \hat{a}_{a2}^\dagger$ and $\hat{a}_{b1}^\dagger \hat{a}_{b2}^\dagger$ do not afford the TPI fringe, because the synchronized phase randomizations by the two PZTs are performed in paths a1 and a2 of the two MZIs. Consequently, the phase-insensitive HOM-type TPI fringe is only observed without the introduction of any relative phase relation between the two amplitudes $\hat{a}_{a1}^\dagger \hat{a}_{b2}^\dagger$ and $\hat{a}_{a2}^\dagger \hat{a}_{b1}^\dagger$. However, if we introduce an additional phase shift $\Delta\phi$ in one of the four paths (for example, path b2), the measured coincidences reveal an oscillation between the HOM peak and dip fringes as a function of the phase shift, as shown in Fig. 10.

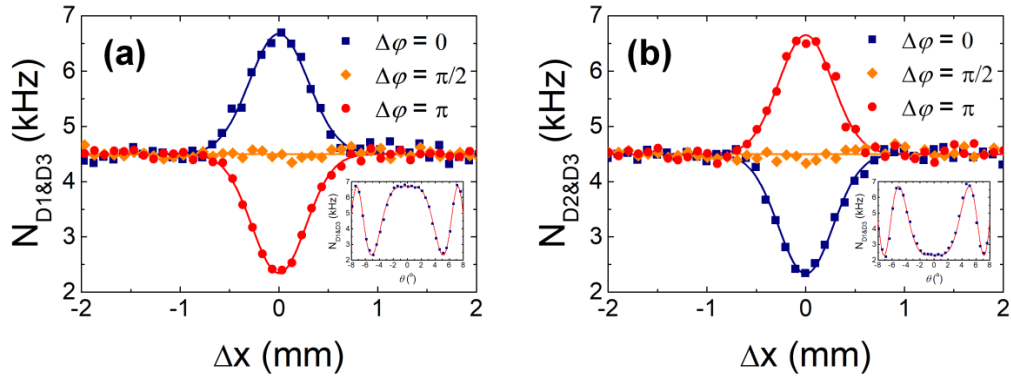

Fig. 10. Experimental results relating to the two-photon interference of spatially separated weak coherent pulses. Coincidences are measured with two detectors (a) D1 and D3, and (b) D2 and D3 as functions of path-length difference $\Delta x = |\Delta x_1 + \Delta x_2|$ for two spatially separated input photons. Coincidence fringes for each combination of the two detectors are obtained with a relative phase shift between the two two-photon amplitudes. Insets in (a) and (b) indicate the coincidence counts measured as functions of the rotation angle of the thin glass plate, which introduces an additional relative phase shift between the two interfering amplitudes.

Next, we consider the case in which the two interfering two-photon amplitudes corresponding to Fig. 9 overlap in the two paths of the single MZI ($a_1 = a_2 = a$, $b_1 = b_2 = b$, and $\Delta x_1 = \Delta x_2 = \Delta x$) shown in Fig. 1(b). In this case, the two-photon amplitudes contributing to coincidence detection by SPDs D1 and D2 should be considered including BS2, that is, $\hat{a}_a^\dagger \hat{a}_b^\dagger \xrightarrow[t-t]{BS2} \hat{a}_{D1}^\dagger \hat{a}_{D2}^\dagger$ and $\hat{a}_a^\dagger \hat{a}_b^\dagger \xrightarrow[r-r]{BS2} \hat{a}_{D1}^\dagger \hat{a}_{D2}^\dagger$, where t and r denote the transmission and reflection of the two photons at BS2, respectively. Indeed, the conventional single MZI scheme corresponds to a folded version of that shown in Fig. 9. As a result, to observe the HOM-type TPI of weak coherent light in two spatially separated MZIs, it is necessary to satisfy the experimental conditions in such a manner that the two-photon amplitudes in a single MZI contribute to the observation of the HOM interference. Although the observed HOM-type TPI fringe with the two spatially separated MZIs can be fully explained by the classical intensity correlation with respect to the limited maximum visibility of 0.5 and the fringe shape, the interference effect including the coincidence-counting rate and the related two-photon state may need to be understood as a consequence of the two-photon correlation at the single-photon level or the interference of two indistinguishable two-photon amplitudes.

**Discussion**

The TPI of highly correlated photon pairs is interpreted as the interference between the two indistinguishable two-photon amplitudes within the utilized interferometers. When the similar experiment is performed by employing weak coherent light at the single photon level, the physics underlying the interference effect needs to be understood in the quantum framework as it originate from the interference of the two indistinguishable two-photon amplitudes at the single-photon level. In this context, we experimentally demonstrated the observation of the TPI of weak coherent pulses and quantitatively analyzed the experimental results considering the single- and coincidence-counting rates based on the statistical property of the coherent state and the interference fringe originating from the two-photon state at the single-photon level. In particular, the HOM-type TPI experiment was performed by employing a phase-randomization mechanism, and further experiments were performed utilizing two temporally- and spatially separated weak coherent pulses.

From the statistical property of the coherent light with a very low mean photon-number $\langle n \rangle \sim 0.03$, we clearly confirmed in the experiments that the observed TPI fringe corresponding to the coincidence-counting events with varying path-length differences originated from the two-photon component in the input pulse. The observed TPI fringe shape and visibility can also be explained by multiplying the two single-photon counting events; in this case, the result is similar to the classical TPI fringe given by the classical intensity correlation function. Moreover, we showed that the HOM-type TPI fringe visibility is limited to 0.5 owing to the path-correlated two-photon state that does not contribute to the interference fringe but only to the random coincidence-counting event caused by the active phase-randomization mechanism.

When the two input photons contributing to the TPI are temporally and spatially well-separated, the observed TPI is more clearly explained by the two two-photon amplitudes formed with separated weak coherent pulses at the single-photon level. According to Dirac's famous statement on single-photon interference[40], and its two-photon analogy[41], if each single-photon or photon-pair interferes only with itself, then individual single photons contributing to the TPI do not necessarily have to travel through a common interferometer with spatiotemporal overlap. In other words, individual photons can separately traverse the interferometer arms to exhibit TPI via the coincidence measurement of two single-photon counting events. For the TPI experiment with weak coherent pulses traveling through the two spatially separated MZIs, when the interferometric scheme is configured in such a manner that the two interfering photons traverse as if through a common interferometer, the observed TPI fringe shows the same feature as the fringe obtained in the single-MZI scheme, except for an additional phase shift introduced into the interfering two-photon states. As a result, the observed TPI fringe shows HOM-type bunching and splitting depending on the relative phase shift between the two two-photon amplitudes contributing to the interference effect.

Although the TPI of weak coherent light, especially the fringe shape and visibility bound, can be described classically by intensity correlation, we emphasize that the observed interference effect should be considered as the interference between two alternative two-photon states constituted in the two interferometer arms when we consider the weak coherent pulse at the single-photon level. We believe that our results can provide more comprehensive understanding of the TPI of weak coherent light at the single-photon level.

**Acknowledgements**

This work was supported by the National Research Foundation of Korea (NRF) (Nos. 2021R1A2B5B03002377, 2020M3E4A1080030 and 2020R1I1A1A01072979) and the Ministry of Science and ICT (MSIT), Korea, under the Information Technology Research Center (ITRC) support program (IITP-2021-2020-0-01606) supervised by the Institute of Information & Communications Technology Planning & Evaluation (IITP). This work was also supported in part under the research program (Nos. 2019-106 and 2020-081) by the affiliated institute of ETRI.

**Author Contributions**

H.K., O.K. and H.S.M. conceived the project. H.K. designed the experimental setup and performed the experiments. H.K., O.K. and H.S.M. discussed the results and contributed to the writing of the manuscript.

**Additional Information**

**Competing Interests:** The authors declare no competing interests.